\documentclass[aps,jcp,twocolumn,groupedaddress,longbibliography]{revtex4-1}
\usepackage[dvipsnames]{xcolor}
\usepackage{graphicx,amsmath,amssymb,color}
\newcommand{\rhedit}[1]{\textcolor{ForestGreen}{#1}}

\begin{document}

\title{Quantitative relations between nearest-neighbor persistence and slow heterogeneous dynamics in supercooled liquids} 
\author{Katrianna S. Sarkar$^1$, Kevin A. Interiano-Alberto$^1$, Jack F. Douglas$^2$, and Robert S. Hoy$^1$}
\affiliation{$^1$Department of Physics, University of South Florida, Tampa, Florida 33620, USA}
\affiliation{$^2$National Institute of Standards and Technology, Gaithersburg, Maryland 20899, USA}
\email{rshoy@usf.edu}
\date{\today}
\begin{abstract}
Using molecular dynamics simulations of a binary Lennard-Jones model of glass-forming liquids, we examine how the decay of the normalized neighbor-persistence function $C_{\rm B}(t)$, which decays from unity at short times to zero at long times as particles lose the neighbors that were present in their original first coordination shell,  compares with those of other, more conventionally utilized relaxation metrics.  
In the strongly-non-Arrhenius temperature regime below the onset temperature $T_{\rm A}$, we find that $C_{\rm B}(t)$ can be described using the same stretched-exponential  functional form  that is often utilized to fit the self-intermediate scattering function $S(q, t)$ of glass-forming liquids in this regime.
The ratio of the bond lifetime $\tau_{\rm bond}$ associated with the terminal decay of $C_{\rm B}(t)$ to the $\alpha$-relaxation time $\tau_\alpha$ varies appreciably and non-monotonically with $T$, peaking at  $\tau_{\rm bond}/\tau_\alpha \simeq 45$ at $T \simeq T_{\rm x}$, where  $T_{\rm x}$ is a crossover temperature separating the high- and low-temperature regimes of glass-formation.
In contrast,  $\tau_{\rm bond}$ remains on the order of the overlap time $\tau_{\rm ov}$ (the time interval over which a typical particle moves by half its diameter), and the peak time $\tau_\chi$ for the susceptibility $\chi_{\rm B}(t)$ associated with the spatial heterogeneity of $C_{\rm B}(t)$ remains on the order of $\tau_{\rm imm}$ (the characteristic lifetime of immobile-particle clusters), even as each of these quantities varies by roughly $5$ orders of magnitude over our studied range of $T$.
Thus, we show that $C_{\rm B}(t)$ and $\chi_{\rm B}(t)$ provide semi-quantitative spatially-averaged measures of the slow heterogeneous dynamics associated with the  persistence of immobile-particle clusters. 
\end{abstract}
\maketitle

\section{Introduction}
\label{sec:intro}

 Numerous experimental and computational studies have shown that glass-forming liquids are dynamically heterogeneous in the sense that they possess transient regions where the relaxation dynamics can be substantially faster or slower than their spatial average  \cite{ediger96,ediger00,sillescu99,cicerone95,heuer95,kegel00,kob97,donati98,donati99,lacevic03,starr13,berthier11book,paeng15,richert15}.
High-mobility particles tend to form clusters with lifetimes scaling as $\tau_{\rm mob} \sim T/D$, where $T$ is temperature and $D$ is the diffusion constant.
Low-mobility  particles tend to form clusters with lifetimes $\tau_{\rm imm} \sim \tau_\alpha$, where $\tau_\alpha$ is  the $\alpha$-relaxation time.
Most studies of dynamical heterogeneity have focused on the mobile-particle clusters and the associated cooperative motion of the mobile particles in relation to understanding  glass-forming liquids' temperature-dependent activation energy $E_{\rm A}(T)$.

Immobile-particle clusters have also been shown  to play a significant role in these liquids' dramatic dynamical slowdown. 
Lacevic \textit{et al.}\ connected these clusters directly to $\alpha$ relaxation more than twenty years ago \cite{lacevic03} by showing that they dominate the four-point position-time correlation function $\chi_4(t)$, and that the time $\tau_{\rm 4}$ at which $\chi_4(t)$ is maximized tracks $\tau_\alpha$ with decreasing $T$.
Many studies performed since then have \textit{defined} ``dynamical heterogeneity''  (DH) in terms of $\chi_4(t)$. 
DH has also been often defined in terms of the more easily calculated and experimentally determined peak in the non-Gaussian parameter $\alpha_2(t)$.
Notably, the time $t^*$ at which $\alpha_2$ is maximized is a ``slow $\beta$'' relaxation time in the sense that it is much longer than the fast $\beta$ relaxation time $\tau_\beta$ governing the initial decay of $S(q,t)$, but much shorter than $\tau_\alpha$, in glass-forming liquids at low temperature\rhedit{s}. 
Recent work has shown that $t^*$ closely tracks that of the Johari-Goldstein relaxation time $\tau_{\rm JG}$ \cite{puosi21,xu22}.

Starr \textit{et al.}\ showed \cite{starr13} that there are two distinct types of dynamic heterogeneity on these time scales [i.e., $\mathcal{O}(t^*)$ and $\mathcal{O}(\tau_\alpha)$] which are respectively associated with the mobile- and immobile-particle clusters; below, we refer to these as ``fast'' and ``slow'' DH.
However, despite the advances made by a few more-recent studies \cite{wang19,zhang19,giuntoli20,gao25}, immobile-particle clusters and slow DH remain relatively poorly studied in comparison to their mobile-particle and fast counterparts, and there is clearly a need to better understand these phenomena.

In this paper, we take a different approach to characterizing the spatially-heterogeneous dynamics associated with the immobile-particle clusters.
In particular, we focus on the neighbor-persistence metric $C_{\rm B}(t)$, which varies smoothly from 1 to 0 as particles lose the neighbors that were present in their original first coordination shell, and captures the extent to which individual particles progressively “forget” their original local environments.  
$C_{\rm B}(t)$ was suggested to be a natural measure of structural relaxation in supercooled liquids as far back as the work of Frenkel in the 1930s \cite{frenkel46}, and Rahman examined related quantities in some of the first molecular dynamics simulations \cite{rahman66}, yet this metric has received surprisingly little attention over the intervening decades.
This neglect can no doubt be traced to the fact that it cannot yet be experimentally measured in atomic and molecular glass-formers.

Experimental measurements of $C_{\rm B}(t)$ have, however, been performed  for \textit{colloidal} glass-formers.
Conrad \textit{et al.}\ showed that their solid-like, elastic response to an applied shear strain on timescales $t < \tau_{\rm solid}$  arises from the system-spanning nature of the immobile-particle clusters on these timescales  \cite{conrad06}.
They showed that these clusters have $C_B(t < \tau_{\rm solid}) = 1$, and further that in the glassy state, such clusters persist throughout the experimental observation window.
Zhang \textit{et al.}\ compared $C_{\rm B}(t)$ in attractive and repulsive glasses composed of the same particles at the same packing fractions \cite{zhang11}, and showed that the much-slower dynamics of the former systems is directly associated with their much-slower-decaying $C_{\rm B}(t)$.
Laurati \textit{et al.}\ extended the results of Ref.\ \cite{conrad06} to mechanically-deformed systems, showing \cite{laurati17} that the shape of the stress-strain curves $\sigma(\gamma)$ under startup shear is intimately connected to $C_{\rm B}(\gamma)$, i.e.\rhedit{,} to the fraction of particles' neighbors in the initial undeformed state which remain their neighbors at shear strain $\gamma$.
Most recently, Higler \textit{et al.}\  measured $C_{\rm B}(t)$ in charge-stabilized colloidal suspensions with a wide range of $\phi$, $\tau_\alpha$, $\tau_{\rm bond}$ and coordination numbers $Z$ \cite{higler18}.
They used these results to reformulate Dyre's shoving model \cite{dyre96,dyre06} for the relationship between $\tau_\alpha$ and the cage-escape lifetime $\tau_{\rm cage}$ (which is of order $\tau_{\rm bond}$) in terms of more-readily-observable particle-scale quantities.

Several additional insights have come from molecular dynamics simulations.
Yamamoto and Onuki showed \cite{yamamoto98} that the time scale $\tau_{\rm bond}$ associated with the terminal decay of $C_{\rm B}(t)$ seemed to satisfy the relation $\tau_{\rm bond} \simeq 10\tau_\alpha$, that reneighboring events become increasingly spatially heterogeneous with decreasing $T$, that $\tau_{\rm bond} \propto \xi^z$ where $\xi$ is the correlation length of such heterogeneities (i.e.\rhedit{,} the size of clusters formed by particles with few lost neighbors), and that the character of these heterogeneities is strongly affected by applied shear.
Shiba \textit{et al.}\ substantially extended these results, showing \cite{shiba12} that $\tau_{\rm 4,pos}$ was substantially shorter than the peak time $\tau_{\rm 4,bond}$ for  four-point bond-breakage correlations, and claimed that this is associated with long-wavelength vibrational modes which reduce the former but not the latter.  
Iwashita \textit{et al.}\ argued  \cite{iwashita13} that $C_{\rm B}(t)$ is directly connected to the super-Arrhenius temperature dependence of fragile glassformers' relaxation dynamics.
In particular, they showed that the Maxwell time ($\tau_{\rm M}$) exceeds the average time for particles to lose \textit{one} neighbor ($\tau_{\rm LC}$) for temperatures below 
$T_{\rm c}$, thus identifying these single-reneighboring events as the elementary excitations in high-temperature and weakly-supercooled liquids.
Most recently, Scalliet \textit{et al.}\ compared the spatially-resolved $C_{\rm B}[\tau_{\rm bond}(T)]$ of supercooled liquids over a wide range of $T$ \cite{scalliet22}.
They showed that the spatial heterogeneity of neighbor persistence increases substantially as $T$ decreases into a regime that is more deeply supercooled than that probed by most previous simulations.
In this regime, structural relaxation corresponds to transient but long-lived solid-like regions with low particle mobility and high $C_{\rm B}$ being gradually ``dissolved'' by liquid-like regions displaying opposite trends.
This interpretation seems to agree with a wide range of recent theoretical work \cite{dyre24}.

Taken together, these results indicate that $C_{\rm B}(t)$ is an appealing relaxation metric to study because it is both experimentally accessible and simple to interpet, yet it can offer deep insights not obtainable via more-commonly studied metrics like particles' mean-squared displacement $\Delta^2(t)$ and self-intermediate scattering function $S(q,t)$.
However, the \textit{quantitative} relationships between the relaxation times associated with $C_{\rm B}(t)$ and several other, more-conventionally-utilized relaxation metrics  have yet to be established.
In this paper, using molecular dynamics simulations of the Kob-Andersen (KA) model \cite{kob94,kob95b}, a  model glass-former which has been intensively studied for three decades, we do so.

We find that for temperatures below $T_{\rm A}$, $C_{\rm B}(t)$ is well fit by the same generic double-stretched-exponential functional form
\begin{equation}
\small C_{\rm B}(t)  =  (1-A) \exp[-(t/\tau_{\rm fast})^{\rm s_{fast}}] + A  \exp[(-t/\tau_{\rm slow})^{\rm s_{slow}}]
\label{eq:commonform}
\end{equation}
that is often utilized to fit  $S(q, t)$ \cite{giuntoli20}.  
The $\tau_{\rm fast}$, $s_{\rm fast}$ and $s_{\rm slow}$ for $C_{\rm B}(t)$ are close to those obtained by fitting the corresponding fast-$\beta$ and $\alpha$ relaxation processes of $S(q,t)$, but the $A(T)$ are somewhat larger.
Consistent with previous studies  \cite{yamamoto98,conrad06,zhang11,shiba12,iwashita13,laurati17,higler18,scalliet22}, the characteristic neighbor lifetime $\tau_{\rm slow}$ is substantially longer than $\tau_\alpha$.
We also find that the ratio $\tau_{\rm slow}/\tau_\alpha$ varies substantially with $T$, and peaks near $T_{\rm x}$.
On the other hand,  $\tau_{\rm slow}$ remains within a factor of order one of $\tau_{\rm ov}$, and $\tau_\chi$ remains within a factor of order one of $\tau_{\rm imm}$, even as each of these quantities varies by roughly $5$ orders of magnitude over our studied range of $T$.
Van Hove correlation functions evaluated at $t = \tau_\chi(T)$ reveal that diffusion remains strongly non-Gaussian on this timescale -- increasingly so as $T$ decreases -- owing to the persistence of large immobile-particle clusters.
Thus we show that $C_{\rm B}(t)$ and $\chi_{\rm B}(t)$ provide easily-measurable, semi-quantitative, spatially-averaged measures of the slow heterogeneous dynamics associated with immobile particles.

\section{Methods}
\label{sec:methods}

All simulations were conducted using the High Dimensional Molecular Dynamics (\texttt{hdMD}) code \cite{hoy22,hdMD}.
To capture generic supercooled-liquid behavior, we simulate the standard Kob-Andersen model \cite{kob94} with a 2:1 ratio of large (A) to small (B) particles.
Particles $i$ and $j$ interact via the truncated and shifted Lennard-Jones potential $U(r) = U_{\rm LJ}(r) - U_{\rm LJ}(r_{\rm c})$, where
\begin{equation}
U_{\rm LJ}(r) = 4\epsilon_{ij} \left[ \left( \displaystyle\frac{\sigma_{ij}}{r} \right)^{12} -  \left(  \displaystyle\frac{\sigma_{ij}}{r} \right)^6  \right]. 
\label{eq:LJpot}
\end{equation}
As usual  \cite{kob94,kob95b}, $\epsilon_{\rm AA} = 1.0$, $\epsilon_{\rm AB} = 1.5$, $\epsilon_{\rm BB} = 0.5$, $\sigma_{\rm AA} = 1.0$, $\sigma_{\rm AB} = 0.8$ and $\sigma_{\rm BB} = 0.88$ in units of the system's characteristic energy and length scales, and the cutoff radius $r_c = 2.5\sigma_{\rm ij}$.

We employ the 2:1 variant of this model rather than the historically-more-widely-studied 4:1 variant because the latter is prone to crystallization \cite{ingebrigtsen19}.
All results reported below will be given in units of $\epsilon_{\rm AA} $, $\sigma_{\rm AA}$ and the time scale $\tau = \sqrt{m\sigma_{\rm AA}^2/\epsilon_{\rm AA}}$, where all particles have mass $m$.
The MD timestep we employed is $dt = \tau/125$.
All systems contain $N = 1.67\times 10^5$ particles and are equilibrated at various constant temperatures $0.38 \leq T \leq 1.0$ and a small positive pressure $P = 0.01$.
In real units corresponding to liquid argon, this pressure corresponds to $P \simeq 420\rm{KPa}$, while our highest temperature  corresponds to $T \simeq 120\rm{K}^\circ$ \cite{sastry01,yuan24}.
As is standard practice \cite{kob95b}, we equilibrate systems for at least 100$\tau_\alpha$ [defined using the standard criterion $S(q, \tau_\alpha) = 1/e$] for each $T$ before beginning the measurements reported below.

Following Refs.\ \cite{yamamoto98,conrad06,zhang11,iwashita13,shiba12,laurati17,higler18,scalliet22}, we define $\mathcal{C}_{B}^i(t)$ as the average fraction of neighbors present in particle $i$'s original first coordination shell that are still present at time $t$.
Then $C_{\rm B}(t) = \langle \mathcal{C}_{B}^i(t) \rangle$ is the average of this quantity over all particles, and the susceptibility $\chi_{\rm B}(t) = \sqrt{\langle [\mathcal{C}^i_{B}(t)]^2 \rangle - C_{\rm B}(t)^2}$ captures the heterogeneity of this metric. 
Most previous studies of this quantity have defined  $\mathcal{C}_{B}^i(t)$ as the average fraction of particle pairs with $r_{ij}(0) \leq B_1 \sigma_{ij}$ that have $r_{ij}(t) \leq B_2 \sigma_{ij}$, where $B_1$ was set to approximately correspond to  the first minimum of the pair correlation functions $g_{ij}(r)$, and $B_2 \simeq 1.1B_1$.
Here, particles $i$ and $j$ are considered neighbors at time $t$ if  $r_{ij}(t) \leq (5/4)2^{1/6}\sigma_{ij} \simeq 1.4\sigma_{ij}$ for \textit{arbitrary} $t$; this effectively sets $B_2 = B_1$.

We will compare the relaxation captured by $C_{\rm B}(t)$ and $\chi_{\rm B}(t)$ to several other more-conventionally-utilized relaxation metrics, specifically:
\begin{itemize}
\item the mean-squared displacement $\Delta^2(t) = \langle |\vec{r}_i(t) - \vec{r}_i(0)|^2 \rangle$ and non-Gaussian parameter $\alpha_2(t) = 3\langle |\vec{r}_i(t) - \vec{r}_i(0)|^4 \rangle/(5[\Delta^2(t)]^2) - 1$ \cite{kob95b,ediger00}; 
\item the self-intermediate scattering function $S(q,t)$, evaluated for $q$ values that are within $0.1/\sigma_{\rm AA}$ of the peak of the static structure factor $S(q)$;
\item the overlap function $f_{\rm ov}(t)$ \cite{lacevic03}, here defined as the fraction of particles that remain within $\sigma_{ii}/2$ of their initial positions;
\item the average sizes $N_{\rm mob} (t)$ and  $N_{\rm imm} (t)$ of mobile- and immobile-particle clusters \cite{kob97,donati98,donati99,lacevic03,starr13}; and
\item the self part of the van Hove correlation function $G_s(r,t) =  N^{-1} \sum_{i = 1}^N \delta\left( |\vec{r}_i(t) - \vec{r}_i(0) | - r \right)$ \cite{kob95b}, where $\delta$(x) is the Dirac delta function.
\end{itemize}

To obtain a relatively simple theoretical picture, it is necessary to extract characteristic timescales from the above relaxation metrics.
Here, following previous studies \cite{kob97,donati98,donati99,lacevic03,starr13}, we define the times $t^*$, $\tau_{\rm mob}$, and $\tau_{\rm imm}$ as the peak times for $\alpha_2(t)$, $N_{\rm mob} (t)$,  and $N_{\rm imm} (t)$, and similarly, we define the time $\tau_\chi$ as the peak time for $\chi_{\rm B}(t)$.
Results for all of these quantities are time-averaged over at least ten ``windows'' of length $10^4\tau$ for $T > 0.42$, four windows of length $10^5\tau$ ($2.5\times 10^5\tau$) for $0.41 \leq T \leq 0.42$ ($0.39 \leq T \leq 0.40$), and two windows of length $5\times 10^5\tau$ for $T = 0.38$.

Rigorously characterizing particles as ``mobile'' or ``immobile'' is a non-trivial exercise.
Here we defined the mobile (immobile) particles as the 10\% of particles that had the largest (smallest) maximal excursions from their initial positions over the time interval $[0,t]$.   
 As in Ref.\ \cite{wang21}, this percentage was chosen because it maximized the peak \textit{normalized} average mobile-particle cluster size, i.e.\ the maximal 
 $N_{\rm mob}(t)/N_{\rm rand}$, where $N_{\rm rand}$ is the average size of clusters formed by randomly selected particles. 

Note that the $f_{\rm ov}(t)$ defined above is the \textit{self}-overlap function, i.e.
\begin{equation}
f^{\rm self}_{\rm ov}(t) = \displaystyle\frac{1}{N} \displaystyle\sum_{i = 1}^N \Theta\left[\sigma_{ii}/2 - |\vec{r}_i(t) - \vec{r}_i(0)|\right],
\label{eq:fovself}
\end{equation}
where $\Theta$ is the Heaviside step function.
Ref.\ \cite{lacevic03} showed that the susceptibility $\chi_4(t) = \langle [f^{\rm mut}_{\rm ov}(t)]^2 \rangle - \langle f^{\rm mut}_{\rm ov}(t) \rangle^2$ of the  \textit{mutual} overlap function $f^{\rm mut}_{\rm ov}(t) = Q/N$, where
\begin{equation}
Q =  \displaystyle\frac{1}{N} \displaystyle\sum_{i = 1}^N \displaystyle\sum_{j= 1}^N  \Theta\left[a\sigma_{ij} - |\vec{r}_j(t) - \vec{r}_i(0)|\right]
\label{eq:fovmut}
\end{equation}
is the integral of the four-point position/time correlation function $\mathcal{G}[\vec{r}_i(0), \vec{r}_j(t), \vec{r}_k(0), \vec{r}_l(t)]$ over all $[\vec{r}_i(0), \vec{r}_j(t), \vec{r}_k(0), \vec{r}_l(t)]$.
It also showed that  $f^{\rm self}_{\rm ov}(t)$ provides the dominant contribution to $f^{\rm mut}_{\rm ov}(t)$, and that choosing $a = 0.3$ maximizes the heights $\chi_4^{\rm max}(T)$ of the peaks of  $\chi_4(t)$.
Many studies which have used this cutoff have \textit{defined} DH in terms of $\chi_4(t)$.
Moreover, it has been argued that $\chi_4^{\rm max}(T)$ scales with the characteristic correlation lengths $\xi_4(T)$ of DH over time scales $t \sim \tau_\alpha$ as  $\chi_4^{\rm max} \sim \xi_4^{2-\varepsilon}$, where the scaling exponent  is material-dependent \cite{karmakar09,karmakar10,flenner10,kim13}.
Below, we will show that our alternative choice (i.e.\ $a = 1/2$) allows us to more effectively characterize systems' slow DH on time scales $t \gg \tau_\alpha$.

\section{Results}
\label{sec:results}

\subsection{Qualitative characterization of eight different structural-relaxation metrics}
\label{subsec:IIIA}

\begin{figure}[h!]
\includegraphics[width=2.83in]{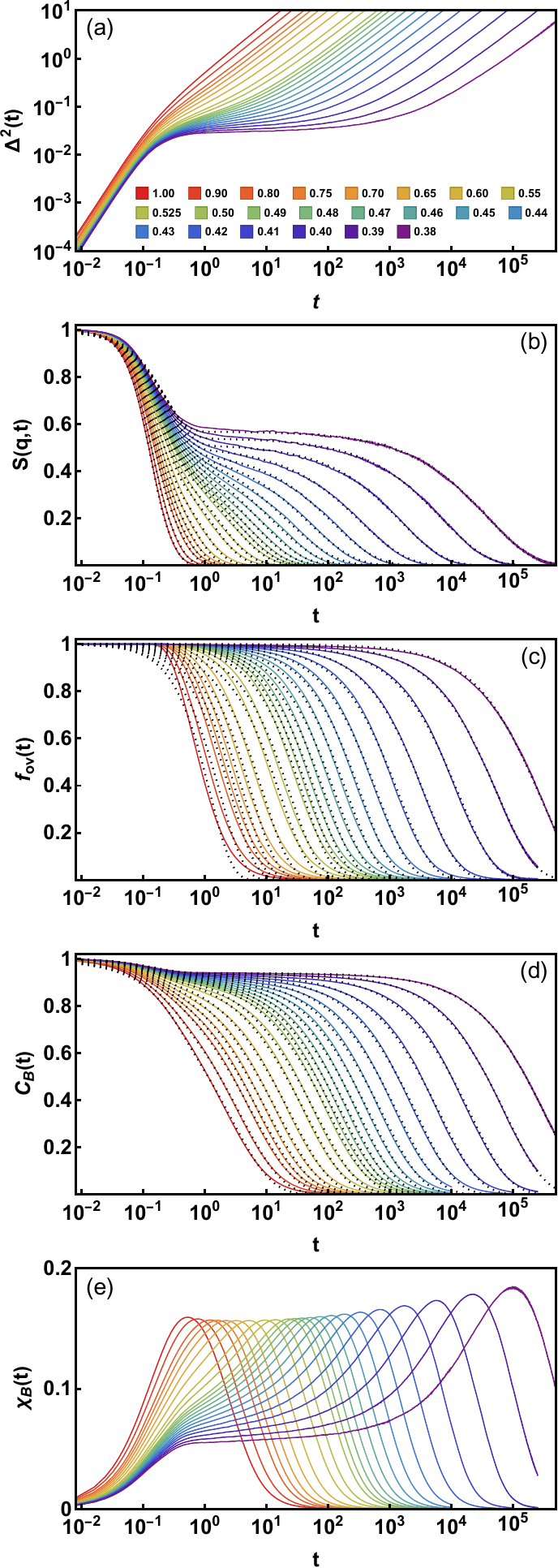}
\caption{Three traditional measures of single-particle dynamics compared to two measures of nearest-neighbor persistence.  Panels (a-e) respectively show $\Delta^2(t)$, $S(q,t)$, $f_{\rm ov}(t)$, $C_{\rm B}(t)$ and $\chi_{\rm B} (t)$ for systems over the temperature range $0.38 \leq T \leq 1.0$.
The color legend in panel (a) shows each $T$ value.
Dotted curves in panels (b-d) show fits to Eqs. \ref{eq:cfSQ}-\ref{eq:cfCB}.}
\label{fig:1}
\end{figure}

Figure \ref{fig:1} summarizes the $T$-  and $t$-dependence for five of the metrics discussed above.
All results shown in panels 1a-1b are similar to those reported in many previous studies \cite{ediger96,kob95b}, but since they set the stage for what will follow, we discuss them in some detail here.
Plateaus in both $\Delta^2(t)$ and $S(q,t)$ begin to form as $T$ drops below the onset temperature for non-Arrhenius relaxation, $T_{\rm A} \simeq 0.64$; this corresponds to the emergence of a distinct fast $\beta$ relaxation process with a nearly $T$-independent relaxation time $\tau_\beta$ and a slow $\alpha$ relaxation process with a strongly $T$-dependent $\tau_\alpha$.
The separation between these timescales upon further cooling exhibits a sharp increase as $T$ drops below $T_{\rm x} \simeq 0.45$.
We will discuss how these estimates of $T_{\rm A}$ and $T_{\rm x}$ were obtained in Section \ref{subsec:IIIB}.

There are two notable differences between the behavior of $\Delta^2(t)$ and $S(q,t)$ and that of $f_{\rm ov}(t)$ (panel 1c).
First, unlike the former metrics, $f_{\rm ov}(t)$ exhibits single-step decay for all $T$.
Second, in contrast to previous studies which employed $a = 0.3$ (Sec.\ \ref{sec:methods}) and found 
that the characteristic relaxation time $\tau_{\rm ov}$ defined by $f_{\rm ov}(\tau_{\rm ov}) = 1/e$ was of order $\tau_\alpha$, the plateaus in  $f_{\rm ov}(t)$ shown here extend to substantially longer times, indicating a $\tau_{\rm ov} \gg \tau_\alpha$.
In other words, throughout the plateau regimes suggested by measurement of particles' mean-squared displacement and self-intermediate scattering function, very few particles have moved by even half their diameter.
This is not surprising given that $r_{\rm cage} \ll \sigma_{ii}/2$, but it does raise the question of how to visualize the physical relaxation process associated with these larger $\tau_{\rm ov}$.

To begin answering this question, we now look at these systems' $C_{\rm B}(t)$ (panel 1d).
At high and intermediate $T$, the $C_{\rm B}(t)$ exhibit smooth, apparently-single-time-scale decays similar to those reported in previous studies  \cite{yamamoto98,conrad06,zhang11,iwashita13,shiba12,laurati17,higler18,scalliet22}.
For $T \lesssim T_{\rm x}$, however, the $C_{\rm B}(t)$ exhibit plateaus which lengthen rapidly with decreasing $T$, much like those for $\Delta^2(t)$, $S(q,t)$ and $f_{\rm ov}(t)$\rhedit{.}
These plateaus have not been previously reported because Refs.\ \cite{yamamoto98,conrad06,zhang11,iwashita13,shiba12,laurati17,higler18,scalliet22} employed $B_2$ which were well above $B_1$ (Sec.\ \ref{sec:methods}), whereas here we employ $B_2 = B_1$.
This difference introduces an additional relaxation mechanism in our $C_{\rm B}(t)$ that was not captured in previous studies:  specifically, particle-pair distances evolving back and forth across the above-mentioned $r_{ij} \leq (5/4)2^{1/6}\sigma_{ij}$ neighboring cutoff.
Below, we will argue that this mechanism is associated with fast $\beta$-relaxation.

Beyond their plateau regimes, the decays of $C_{\rm B}(t)$ and $f_{\rm ov}(t)$ closely track each other, indicating that particles losing their neighbors roughly corresponds to them moving by more than half their diameters.
This is not surprising from a qualitative point of view,  but as we will show below, it leads to an interpretation of the connection of $C_{\rm B}(t)$ to other measures of heterogeneous relaxation dynamics that is substantially different from those developed in previous studies which employed $a = 0.3$ \cite{yamamoto98,iwashita13,lacevic03,shiba12,scalliet22}.
Note that panels 1b-1d also show fits of $S(q,t)$, $f_{\rm ov}(t)$, and $C_{\rm B}(t)$ to Eq.\ \ref{eq:commonform}, but before discussing these, we will turn our attention to measures of heterogeneous dynamics in these systems.

At the same high $T$ for which $S(q,t)$ and $C_{\rm B}(t)$ exhibit single-time-scale relaxation, $\chi_{\rm B}(t)$ takes on an approximately log-normal form (panel 1e).
As $T$ drops below $T_{\rm x}$, clear plateaus of $\chi_{\rm B}(t)$ that mirror those of $C_{\rm B}(t)$ emerge.
Beyond the plateau regime, the heights $\chi_{\rm B}^{\rm max}(T)$ of the peaks in the $\chi_{\rm B}(t)$ curves begin increasing with decreasing $T$.
Note that the gradual emergence of two-time-scale decay of  $S(q,t)$ and $C_{\rm B}(t)$\rhedit{,} as $T$ decreases\rhedit{,} coincides with (1) a decrease in the maximal slopes $d\chi_{\rm B}/d[\ln(t)]$, (2) a reduction of the $\chi_{\rm B}$ values at which the inflection points in the $\chi_{\rm B}(t)$ curves occur, and (3) a rapid increase of $\tau_\chi$.
As we will discuss further below, all of these trends indicate that $C_{\rm B}(t)$ and $\chi_{\rm B}(t)$ as defined in this paper can serve as useful \textit{spatially-averaged} measures of the slow DH associated with the relatively immobile particles.

\begin{figure}[htbp]
\includegraphics[width=2.95in]{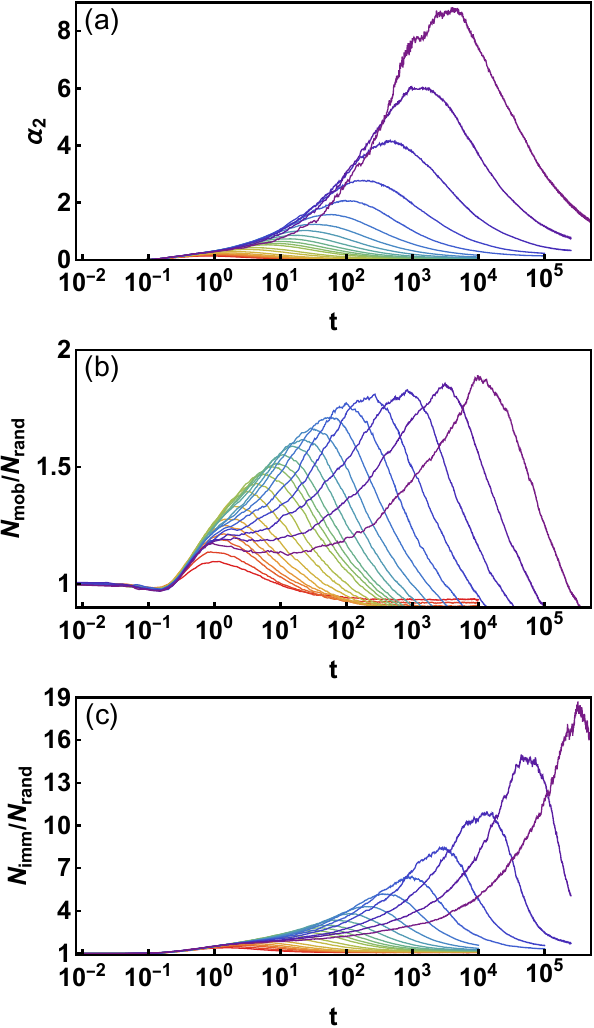}
\caption{Traditional measures of dynamical heterogeneity.  Panels (a-c) respectively show $\alpha_2(t)$, $N_{\rm mob}(t)$, and $N_{\rm imm}(t)$ for all systems; colors are the same as in Fig.\ \ref{fig:1}.}
\label{fig:2}
\end{figure}

Figure \ref{fig:2} shows results for three more metrics which are often employed to characterize dynamical heterogeneity.
In all panels, the trends with decreasing $T$ are consistent with many previous studies \cite{kob97,donati98,donati99,lacevic03,starr13}.
More specifically, the maximum values of $\alpha_2(t)$, $N_{\rm mob}(t)$, and $N_{\rm imm}(t)$, and  and the times $t^*$,  $\tau_{\rm mob}$, and $\tau_{\rm imm}$ at which these maxima occur each increase rapidly with decreasing $T$ for $T \lesssim T_{\rm x}$.
The increasing separation between fast DH timescales like $t^*$ and $\tau_{\rm mob}$ and slow DH timescales like $\tau_{\rm imm}$ has been discussed extensively \cite{ediger96,sillescu99,ediger00,starr13}.
However, as mentioned in the Introduction, while most previous studies have focused on assigning a physical meaning to the fast timescales (e.g.\rhedit{,} showing that  $t^* \sim \tau_{\rm mob} \sim T/D$ \cite{kob97,donati98,donati99,lacevic03,starr13}), here we will focus on the relatively slow particles.
More specifically, we will quantitatively relate $\tau_{\rm imm}$, which measures the characteristic lifetime of  immobile regions whose characteristic size grows very rapidly with increasing $T$ \cite{lacevic03,starr13}, to \textit{local, single-length-scale} neighbor-persistence timescales obtainable from measurements of $C_{\rm B}(t)$ and $\chi_{\rm B}(T)$.

\subsection{Quantitative comparison of seven different relaxation times}
\label{subsec:IIIB}

Next\rhedit{,} we fit the data shown in Fig.\ \ref{fig:1}(a-c) to the common, non-exponential functional form discussed above (Eq.\ \ref{eq:commonform}).
To avoid ambiguity in the following discussion, we will first rewrite Eq.\ \ref{eq:commonform} in ways that are specific to the three quantities of interest.
For the self-intermediate scattering function, we rewrite it as \cite{simmons11,zhang17b,giuntoli20}
\begin{equation}
S(q,t) =  (1-A_{\rm S}) \exp[-(t/\tau_\beta)^{\rm s_\beta}] + A_{\rm S}  \exp[(-t/\tau_\alpha)^{\rm s_\alpha}],
\label{eq:cfSQ}
\end{equation}
where $A_{\rm S}$ is the well-known ``non-ergodicity'' parameter \cite{mountain93}, $\tau_\beta$ and $\tau_\alpha$ are the fast-$\beta$ and $\alpha$ relaxation times, and $s_\beta$, $s_\alpha$ are the associated  exponents quantifying the degree of deviation from exponential relaxation.
This form should accurately describe $S(q,t)$ for temperatures $T < T_{\rm A}$, where  $\tau_\alpha \gg \tau_\beta$.

For the overlap function,  the fact that $f_{\rm ov}(t) \equiv 1$ until particles begin moving at last half their diameters away from their initial positions allows us to set $A = 1$ and therefore to reduce the number of parameters in  Eq.\ \ref{eq:commonform}  from five to two.
Thus, we rewrite Eq.\ \ref{eq:commonform} as
\begin{equation}
f_{\rm ov}(t) \simeq \exp[(-t/\tau_{\rm ov})^{\rm s_{\rm ov}}],
\label{eq:cfFO}
\end{equation}
where $\tau_{\rm ov}$ is the terminal relaxation time associated with this relaxation process and $s_{\rm ov}$ is the associated stretching exponent.
The poor agreement of this approximate functional form with simulation results for $t \lesssim 10^0$ [Fig.\ \ref{fig:1}(c)] is not a major concern for the purposes of this study because we are focusing on the terminal relaxation.
Note that this issue could have been avoided by using  $f_{\rm ov}^{mut}(t)$ [Eq.\ \ref{eq:fovmut}], which exhibits a two-step decay \cite{lacevic03} and can be fit using the same functional form as Eqs.\ \ref{eq:commonform} and \ref{eq:cfSQ}.

Finally, for the neighbor-persistence function, we rewrite Eq.\ \ref{eq:commonform} as
\begin{equation}
\small
C_{\rm B}(t) =  (1-A_{\rm C}) \exp[-(t/\tau_{\rm fast})^{\rm s_{\rm fast}}] + A_{\rm C}  \exp[(-t/\tau_{\rm bond})^{\rm s_{\rm bond}}]\rhedit{.}
\label{eq:cfCB}
\end{equation}
Here $A_C$ plays a role similar to the ergodicity parameter, the fast-relaxation parameters $\tau_{\rm fast}$ and $s_{\rm fast}$ capture the behavior of $C_{\rm B}(t)$'s abovementioned fast decay mode, and $\tau_{\rm bond}$, $s_{\rm bond}$ are the average nearest-neighbor lifetime and associated stretching exponent discussed in Refs.\  \cite{yamamoto98,shiba12}.
Our best-fit values of the parameters in Eqs.\ \ref{eq:cfSQ}-\ref{eq:cfCB}, along with uncertainty estimates, are given in the Appendix.

Before proceeding further, we define both the onset temperature $T_{\rm A}$ and the crossover temperature $T_{\rm x}$.
These temperatures are typically estimated using the $\alpha$ relaxation times.
$T_{\rm A}$ has traditionally been defined \cite{kob95b} as the temperature below which $\tau_\alpha$ increases super-Arrheniusly with decreasing $T$ and substantial dynamical heterogeneity emerges \cite{sastry98}.
A precise criterion was given in Ref.\ \cite{yuan24}, which showed that defining $T_{\rm A}$ as the highest temperature for which $\tau_\alpha(T)$ is 1\% above its high-$T$ Arrhenius fit value allows $\tau_\alpha(T_{\rm A})$ to be identified with the ``caging onset time'' and to serve as a reference time scale for structural relaxation and diffusion.

 $T_{\rm x}$ has traditionally been defined \cite{kob95b} with the critical temperature $T_{\rm c}$ of mode-coupling theory, by fitting $\tau_{\alpha}(T)$ to its predicted power-law divergence
 $\tau_\alpha(T) \sim (T_{\rm c} - T)^{-\gamma}$ \cite{bengtzelius84}, over a limited temperature range.
Values of $T_{\rm c}$ obtained in this fashion depend strongly on the choice of temperature range over which $\tau_\alpha(T)$ is fit to this functional form \cite{berthier10}, however,  and $\tau_\alpha(T)$ does not actually diverge at the fitted $T_c$.
An alternative estimate of the crossover temperature $T_{\rm x}$ can be obtained by finding the maximum of $d^2[\ln(\tau_\alpha)]/d(1/T)^2$ on a Stickel plot \cite{stickel96}, or (roughly equivalently for systems exhibiting a fragile-to-strong crossover \cite{starr99}\rhedit{)} by finding the intersection of the high-$T$ and low-$T$ Arrhenius fits to $\tau_\alpha(T)$ \cite{casalini05}.
This definition of $T_{\rm x}$ is comparable to the crossover temperature defined in the Generalized Entropy Theory (GET) \cite{dudowicz08} that separates the high- and low-temperature regimes of glass formation.
Note that the GET also predicts a power-law variation of $\tau_\alpha$ over a limited $T$ range and predicts the precise value of this crossover temperature.

\begin{figure}[htbp]
\includegraphics[width=3in]{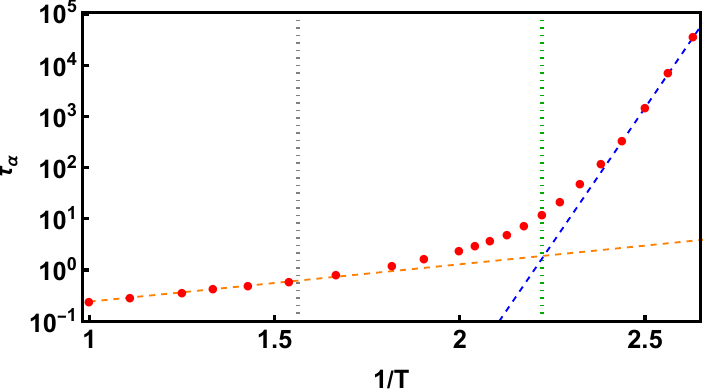}
\caption{$\alpha$ relaxation times  obtained by fitting $S(q,t)$ to Eq.\ \ref{eq:cfSQ}.
Dashed lines show fits of $\tau_\alpha(T)$ to the Arrhenius prediction $\tau_{\rm Arr}(T) = \tau_{\rm 0} \exp(E_{\rm Arr}/T)$; the high- and low-$T$ fits intersect at $T = T_{\rm x} \simeq 0.45$.  
Dotted lines indicate $1/T_{\rm A}$ and $1/T_{\rm x}$.}
\label{fig:3}
\end{figure}

Figure \ref{fig:3} shows the $\tau_\alpha(T)$ obtained from fitting our $S(q,t)$ data to Eq.\ \ref{eq:cfSQ}.
The data are consistent with a previous study of the KA model's 2:1 variant at $P = 0$ \cite{crowther15}, and all trends are similar to those observed in many previous studies.
Here we emphasize two key features.
First, $\tau_\alpha$ increases by more than five orders of magnitude (from $\simeq 0.24$ to $\simeq 3.7 \times 10^4$) as $T$ decreases from $1.0$ to $0.38$.
Second, the procedures described above \cite{yuan24,casalini05} respectively yield $T_{\rm A} \simeq 0.64$ and $T_{\rm x} \simeq 0.45$; for reference, $\tau_\alpha(T_{\rm A}) \simeq 0.62$ and $\tau_\alpha(T_{\rm x}) \simeq 12$.

We will show below that this estimate of $T_{\rm x}$ closely corresponds to two other characteristic features in the temperature dependence of our systems' relaxation dynamics.
We also remark that the characteristic temperatures of the Kob-Andersen model dependent strongly on pressure \cite{yuan24}, and this should be borne in mind when comparing our results to those from the many previous simulations of the Kob-Andersen model that employ constant density.
The values that we find for $T_{\rm A}$ and $T_{\rm x}$ are roughly consistent with corresponding estimates for the 4:1 KA model when $P$ is small.

Detailed results for $A_{\rm S}$, $A_{\rm C}$, $\tau_\beta$, $\tau_{\rm fast}$,  $s_\beta$, $s_\alpha$, $s_{\rm ov}$, $s_{\rm fast}$ and $s_{\rm bond}$ are given in the Appendix.
Here we emphasize that for all temperatures which are low enough for relaxation to have a clearly-two-step character (i.e.\rhedit{,}  $T \leq T_{\rm A}$), $\tau_{\rm fast}$ and $s_{\rm fast}$ are respectively close to $\tau_\beta$ and $s_\beta$.
This shows that  the initial stage of $C_{\rm B}(t)$'s decay is closely associated with the fast $\beta$ relaxation \cite{ediger96}.
We also find that $s_{\rm bond} \simeq s_\alpha$ for $T \lesssim 0.5$, suggesting that neighbor persistence and $\alpha$ relaxation are similarly heterogeneous in this regime.

\begin{figure}[htbp]
\includegraphics[width=3in]{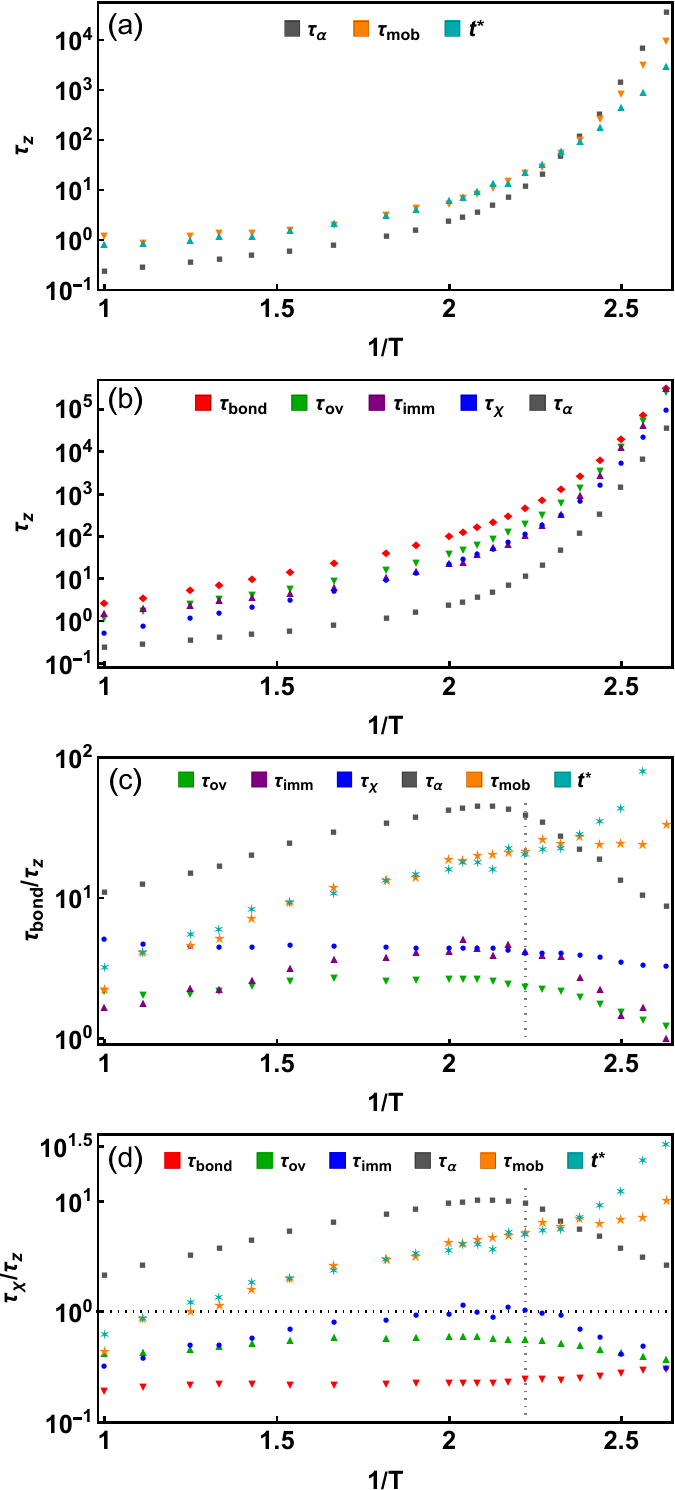}
\caption{ Comparison of seven different relaxation times $\tau_{\rm z}$  in 2:1 KA liquids at $P = .01$.
Panels (a-b) respectively compare the relaxation times associated with $C_{\rm B}(t)$ and $S(q,t)$ to those associated with other metrics for fast and slow DH, while panels (c-d) respectively show the ratios of $\tau_{\rm bond}$ and $\tau_\chi$ to these times.  
Vertical dotted lines in panels (c-d) show $1/T_{\rm x}$, and the horizontal dotted line in panel (d) shows $\tau_\chi/\tau_{\rm z}= 1$. }
\label{fig:4}
\end{figure}

Figure \ref{fig:4} shows results for the seven other relaxation times defined above.
Panel 4a compares the $\tau_\alpha(T)$ obtained by fitting $S(q,t)$ to Eq\rhedit{.}\ \ref{eq:cfSQ} to the two measures of fast DH shown in Fig.\ \ref{fig:2}(a-b).
All results are similar to those reported in multiple previous studies \cite{kob97,donati98,donati99,lacevic03,starr13}.
The main features we wish to highlight here are that: (1) $t^*$ and $\tau_{\rm mob}$ are essentially identical on the scale of the plot, except at the lowest $T$, where $\tau_{\rm mob}$ is clearly the larger of the two; and (2) both of these times grow much more slowly with decreasing $T$ than $\tau_\alpha(T)$.

Panel 4b compares these  $\tau_\alpha(T)$ to $\tau_{\rm imm}$ [Fig.\ \ref{fig:2}(d)], $\tau_\chi$, the $\tau_{\rm ov}$ obtained by fitting $f_{\rm ov}(t)$ to Eq.\  \ref{eq:cfFO}, and the $\tau_{\rm bond}$ obtained by fitting the $C_{\rm B}(t)$ data to Eq.\ \ref{eq:cfCB}. 
Three trends are apparent from examining the relations between these quantities.
First, as previously reported in Refs.\ \cite{yamamoto98,shiba12,iwashita13,scalliet22}, $\tau_\alpha$ remains well below $\tau_{\rm bond}$  for all $T$.
Second, in contrast to the results shown in panel 4a, $\tau_\alpha$ also remains well below the other three timescales for all $T$, and for high and intermediate temperatures,  $\tau_\alpha$ grows slower with decreasing $T$.
Third, $\tau_{\rm bond}$ is substantially larger than $\tau_{\rm ov}$ at high $T$, but these times converge as $T$ decreases, consistent with the geometrically-intuitive notion that particles losing their original neighbors corresponds to them moving by roughly half their diameter away from their initial positions \textit{when caging is strong and motion is hopping-dominated}, i.e.\rhedit{,} at low $T$.

Panels 4c and 4d, which respectively show the ratios of $\tau_{\rm bond}$ and $\tau_\chi$ to the other timescales, illustrate the abovementioned relationships in more detail.
Again, three key trends are apparent.
First, the ratios of $\tau_{\rm bond}$ and $\tau_\chi$ to the fast-DH timescales $t^*$ and $\tau_{\rm mob}$ increase steadily with decreasing $T$.
The ratio of $\tau_{\rm bond}$ to these timescales reaches $\mathcal{O}(10^2)$ at our lowest $T = 0.38$, showing that the much-discussed decoupling \cite{ediger96,ediger00} of the fast and slow relaxation timescales in supercooled liquids also occurs for timescales associated with neighbor persistence.
Second, $0.3 < \tau_\chi/\tau_{\rm imm} < 1.2$ and  $1.0 < \tau_{\rm bond}/\tau_{\rm imm} < 5.2$ over the entire range of $T$, even though all three quantities increase by $5$ orders of magnitude as $T$ decreases from $1.0$ to $0.38$.
This shows that $C_{\rm B}(t)$ and $\chi_{\rm B}(t)$ are closely associated with the immobile-particle clusters.

A third trend evident in these panels is that the ratios $\tau_{\rm bond}/\tau_\alpha$ and $\tau_{\rm bond}/\tau_{\rm imm}$ closely track each other, and both appear to peak at $T \simeq 0.48$, which is only slightly above our estimated $T_{\rm x}$.
Previous studies have established that $\tau_{\rm imm} \sim \tau_\alpha$ \cite{lacevic03,starr13}.
We find that the ratio of these two quantities varies over the narrow range $6.5 < \tau_{\rm imm}/\tau_\alpha < 11.5$ despite the fact that each quantity varies by $5$ orders of magnitude over our studied range of $T$.
The peaks in $\tau_{\rm bond}/\tau_{\rm imm}$ and $\tau_{\rm bond}/\tau_\alpha$ at $T \simeq T_{\rm x}$ point to a secondary effect that is not captured by the abovementioned connections between $\tau_{\rm bond}$, $\tau_{\rm \chi}$ and $\tau_{\rm imm}$.

We believe that these peaks are directly associated with \textit{decoupling}. 
Decoupling in an experimental context is often associated with the ``breakdown'' of the Stokes-Einstein scaling relationship $D \sim \eta^{-1}$ between fluids' mass diffusion coefficient $D$ and shear viscosity $\eta$ \cite{ediger96}.
This breakdown arises when the relatively long timescales associated with the immobile particles and structural relaxation (e.g.,  $\tau_\chi$ and $\tau_\alpha$) start growing faster than the shorter timescales associated with the mobile particles and  the average rate of molecular diffusion (e.g., $\tau_{\rm mob}$ and $t^*$) with decreasing $T$.
It is qualitatively described by  the``fractional'' Stokes Einstein relation \cite{douglas98}, a general power-law scaling relation $t^* \sim \tau_\alpha^{1-\zeta}$ between these characteristic times, where $\zeta < 1$ is the exponent characterizing the strength of decoupling \cite{starr13}.
Recent work \cite{xu22,xu23,yuan24,yuan24b} has also connected decoupling directly to $\tau_\beta$, and related this basic molecular timescale to the much longer timescales $t^*$ and $\tau_\alpha$ which are in turn connected directly to  the decoupling of $D$ and $\eta$.
In particular,  Yuan \textit{et al.}  found that the 4:1 KA model obeys $t^*/\tau_\beta \sim (\tau_\alpha/\tau_\beta)^{1-\zeta}$, with $\zeta \simeq 0.26$, for a wide range of densities and pressures  \cite{yuan24,yuan24b}.

\begin{figure}[htbp]
\includegraphics[width=3in]{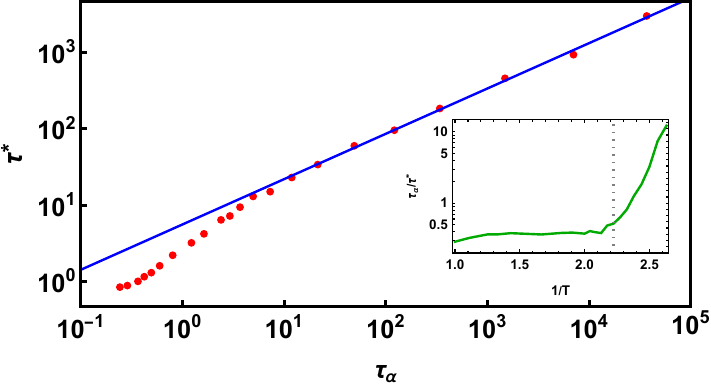}
\caption{Coincidence of the peaks in $\tau_{\rm bond}/\tau_\alpha$ and $\tau_{\rm bond}/\tau_{\rm imm}$ with the onset of strong decoupling.  The solid line shows a fit of the data to $t^* \sim \tau_\alpha^{1-\zeta}$ (with $\zeta = 0.41$).   In the inset, the curve shows $\tau_\alpha/t^*$ for the same dataset shown in Fig.\ \ref{fig:4}, while the vertical dotted line shows $1/T_{\rm x}$.}
\label{fig:5}
\end{figure}

As shown in Figure \ref{fig:5}, both the crossover to $t^* \sim \tau_\alpha^{1-\zeta}$ scaling and the onset of rapidly increasing $\tau_\alpha/t^*$ occur at $T \simeq T_{\rm x} \simeq 0.45$ in our systems.
Thus, the peaks in $\tau_{\rm bond}/\tau_\alpha$  and $\tau_{\rm bond}/\tau_{\rm imm}$, and the subsequent decrease of these quantities with decreasing $T$, can be interpreted as part of the crossover to the low-temperature ($T < T_{\rm x}$) regime of glass-formation \cite{dudowicz08}.
The $\zeta \simeq 0.41$ value reported here is larger than that reported in Ref.\ \cite{yuan24} because our measurements of $\tau_\alpha$ and $t^*$ were obtained from the $S(q,t)$ and $\alpha_2(t)$ of \textit{all} particles, whereas those performed in Ref.\ \cite{yuan24} included the dynamics of $A$ particles only, as is the common practice \cite{kob95b}.
Adopting this practice for our systems decreases $\zeta$ to $\simeq 0.30$ and shifts the crossover to $t^* \sim \tau_\alpha^{1-\zeta}$ scaling upwards, to $T \simeq 0.50$.
Here we prefer to use the values associated with all particles because metrics of heterogeneous dynamics such as $\tau_{\rm bond}$, $\tau_{\rm mob}$ and $\tau_{\rm imm}$ necessarily include all particles.
However, the slightly higher estimate (i.e., $T \simeq 0.50$) is also close to the peaks in  $\tau_{\rm bond}/\tau_\alpha$  and $\tau_{\rm bond}/\tau_{\rm imm}$.
Finally, it must be noted that the $P$ value we have chosen to employ (i.e., $P = 0.01$) is exceptionally low in comparison to most previous studies, and the choice of $P$ apparently influences the point where the power law scaling sets in. This phenomenon requires further study.

\subsection{Connections to heterogeneous caging}
\label{subsec:IIIC}

Next, we connect the above results to  heterogeneous \textit{caging}.
In three spatial dimensions, the probability $P(r,t)$ that a particle has moved a distance $r$ away from its initial position after a time $t$ is $P(r, t) = G_s(r,t)/(4\pi r^2)$. 
According to Einstein's theory of Brownian motion, $P(r,t)$ is Gaussian, and the central limit theorem \textit{requires} that $P(r,t)$ becomes Gaussian after sufficiently long times for a fluid in thermal equilibrium.
At shorter times, however, $P(r, t)$ is clearly non-Gaussian in a wide variety of glass-forming liquids, including particulate systems near their glass and jamming transitions \cite{weeks00,chaudhuri07}.

Exponential tails of form $P_E(r,t) \propto \exp[-r/\Lambda]$, where $\Lambda$ increases slowly with $t$ \cite{chaudhuri07,wang09,wang12,barkai20}, have been claimed to be universal in systems where motion is hopping-dominated \cite{chaudhuri07,barkai20}.
The slow crossovers to Gaussian $P(r,t)$ have been directly associated \cite{wang09,wang12} with the approach to ergodicity.
The exponential tails $P_E(r,t) \propto \exp[-r/\Lambda]$ are known to correspond to the \textit{mobile} particles \cite{chaudhuri07,barkai20}.
In contrast, the connection of non-Gaussian $P(r,t)$ on timescales $t \gg \tau_\alpha$ to \textit{immobile} particles and slow DH remains rather poorly characterized.
In particular, $P(r,t)$ has not yet been directly related to $C_{\rm B}(t)$ or the other measures of slow DH discussed above.
Here we do so.

\begin{figure}[h!]
\includegraphics[width=3in]{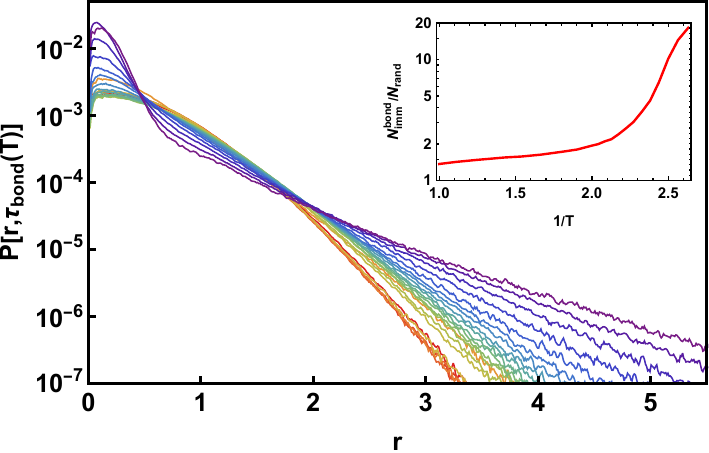}
\caption{Normalized van Hove correlation functions evaluated at the $t = \tau_{\rm bond}(T)$ obtained by fitting $C_{\rm B}(t)$ to Eq.\ \ref{eq:cfCB}.  The inset shows the $N_{\rm imm}[\tau_{\rm bond}(T)]/N_{\rm rand}$ from Fig.\ \ref{fig:2}(c).  Colors are the same as in Fig.\ \ref{fig:1}.}
\label{fig:6}
\end{figure}

Figure \ref{fig:6} shows that $P[r,\tau_{\rm bond}(T)]$ becomes increasingly non-Gaussian as $T$ decreases.
Two notable trends are evident.
First, the lengths of the exponential tails in $P(r)$ increase rapidly with decreasing $T$ for all $T \lesssim 0.7$.
Since these tails correspond to the mobile particles, this trend reflects the increasing $\tau_{\rm bond}/t^*$ and $\tau_{\rm bond}/\tau_{\rm mob}$ shown in Fig.\ \ref{fig:4}(c). 
Second, a prominent low-$r$ peak develops as $T$ drops below $T_{\rm x}$.
This is directly associated with the above-mentioned increasing peak height of $\chi_{\rm B}(t)$ [Fig.\ \ref{fig:1}(e)], and more generally, with the increasing spatial contrast between liquid-like mobile and solid-like immobile regions discussed in Ref.\ \cite{scalliet22}.
The inset shows that it is also directly associated with the persistence of increasingly-large immobile-particle clusters.

Our results indicate that diffusion does not become fully Gaussian (and ergodicity is not recovered) until $t$ is several times larger than $t = \tau_{\rm bond}(T)$.  
While this conclusion is consistent with those of Refs.\ \cite{yamamoto98,conrad06,zhang11,shiba12,iwashita13,laurati17,higler18,scalliet22}, it had not previously been supported so decisively.
Moreover, since  $\tau_{\rm bond} > 10\tau_\alpha$ for all $T$ and indeed  $\tau_{\rm bond} > 40 \tau_\alpha$ for $T \simeq T_{\rm x}$, our results also provide a microscopic justification for the commonly-employed $t_{\rm eq} > 100\tau_\alpha$ criterion for equilibrating supercooled liquids \cite{kob95b}.

\begin{figure}[h!]
\includegraphics[width=3.25in]{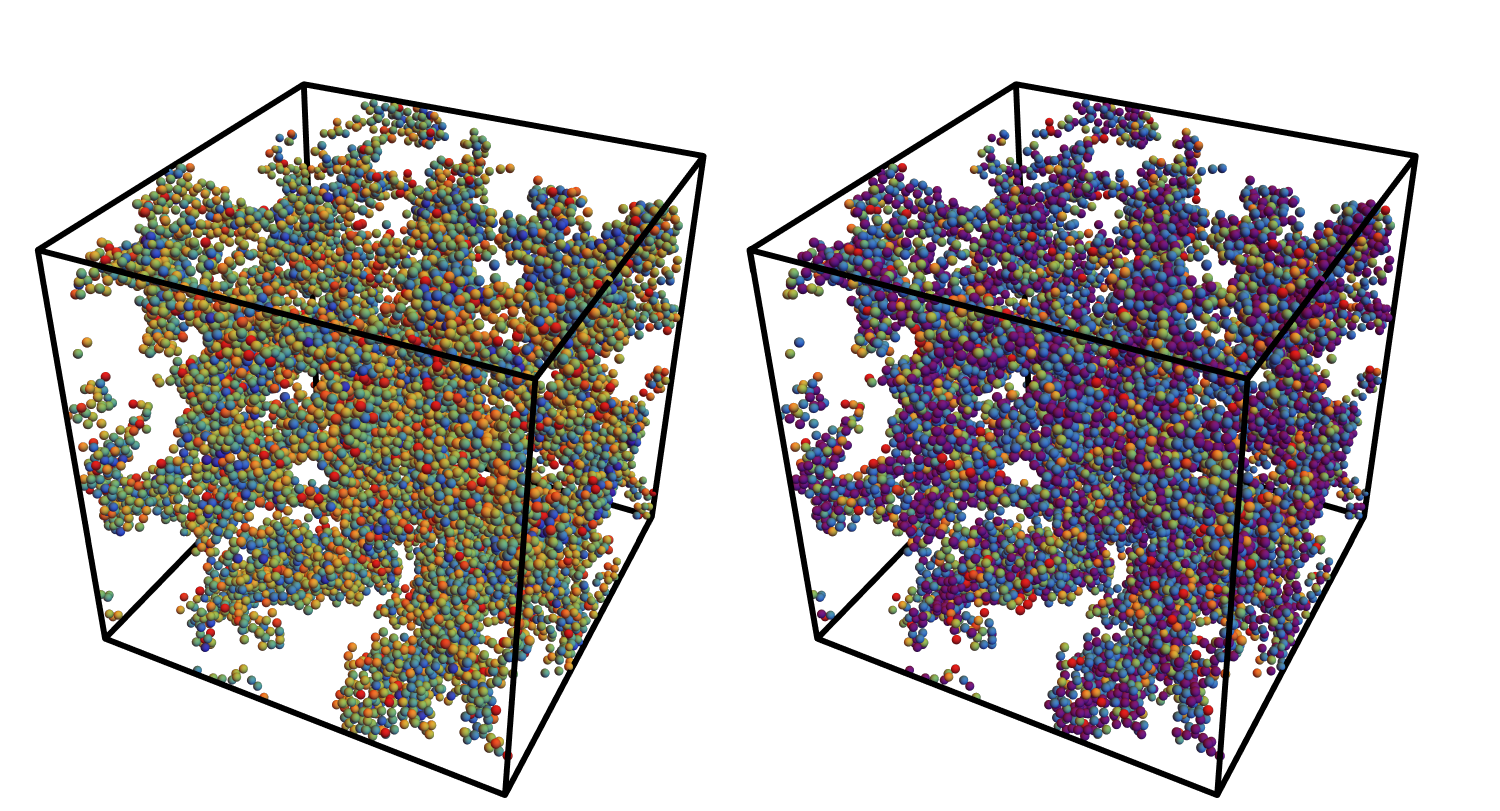}
\caption{Snapshots of the largest (14682-atom) immobile cluster at $t = \tau_\chi(T)$ for $T = \rhedit{0}.38$.  The left image color-codes particles by their maximal displacements $\Delta_{\rm max}$ over the interval $0 < t \leq \tau_\chi$, while the right image color-codes particles by their $C_{\rm B}(\tau_\chi)$.  Colors vary from purple to red for $0 \leq d_{\rm max} \leq 0.3$ and $1 \geq C_{\rm B}(\tau_\chi) \geq 0.75$, respectively. }
\label{fig:7}
\end{figure}

To illustrate the connections between the various trends discussed above, Figure \ref{fig:7} shows snapshots of the largest immobile-particle cluster at $t = \tau_\chi(T)$ for $T = 0.38$.
This cluster contains 14,682 atoms, i.e.\ over 85\% of all the $N/10$ immobile particles present in the system.
It exhibits a fractal structure familiar from previous studies examining the moderately-supercooled regime \cite{starr13,zhang15,wang19,scalliet22}, and percolates along all three directions.
The left image color-codes particles by their \textit{maximal} displacements over the interval $0 < t \leq \tau_\chi$, while the right image color-codes particles by their $C_{\rm B}(\tau_\chi)$. 
The color variations in the two images match up well, i.e.\ the least-mobile particles in the left image appear to correspond closely to the highest-$C_{\rm B}$ particles in the right image.
More specifically, more than 99\% of the atoms in this cluster have $\Delta(\tau_\chi) \equiv \langle | \vec{r}_i(\tau_\chi) - \vec{r}_i(0) | \rangle \leq 0.3$ and $C_{\rm B}(\tau_\chi) \geq 0.75$.
Thus the low-$r$ peak in $P(r,t)$ (Fig.\ \ref{fig:6}) is dominated by the atoms in this cluster.

\section{Discussion and Conclusions}
\label{sec:conclude}

Understanding the dramatic slowdown of supercooled liquids' dynamics with decreasing temperature $T$ is severely hampered by the fact that only a few of the relaxation metrics considered in theories and simulations are readily measurable in experiments.
For example, measurements of the diffusion constant $D$ and alpha relaxation time $\tau_\alpha$ can be used to investigate characteristic aspects of glassy dynamics such as the breakdown of the Stokes-Einstein relation ($D\tau_\alpha = \rm{constant}$) \cite{ediger96,ediger00}\rhedit{.
M}odern probe-molecule-reorientation experiments can provide a great deal of information about dynamical heterogeneity by characterizing how the exponent $\beta$ associated with stretched-exponential relaxation functions of the common form $F(t) \sim \exp[-(t/\tau)^\beta]$ depends on the length and time scales over which relaxation is probed \cite{richert15,paeng15},
However, such phenomenological observational trends do not provide a clear physical understanding of the molecular origin of these phenomena.

Here we have shown that a simple neighbor-persistence metric $C_{\rm B}(t)$, which \textit{is} readily experimentally accessible in colloidal glassformers \cite{conrad06,zhang11,laurati17,higler18}, can help resolve this issue.
In fact, $C_{\rm B}(t)$ and its variance $\chi_{\rm B}(t)$ provide near-quantitative metrics for the slow, spatially-heterogeneous dynamics which occur over timescales from $10$ to $100\tau_\alpha$ and length scales far above those of individual atoms' first coordination shells.
More specifically, $C_{\rm B}(t)$ and $\chi_{\rm B}(t)$ provide spatially-averaged measures of the dynamics associated with immobile-particle clusters.

\begin{figure*}[htbp]
\includegraphics[width=7in]{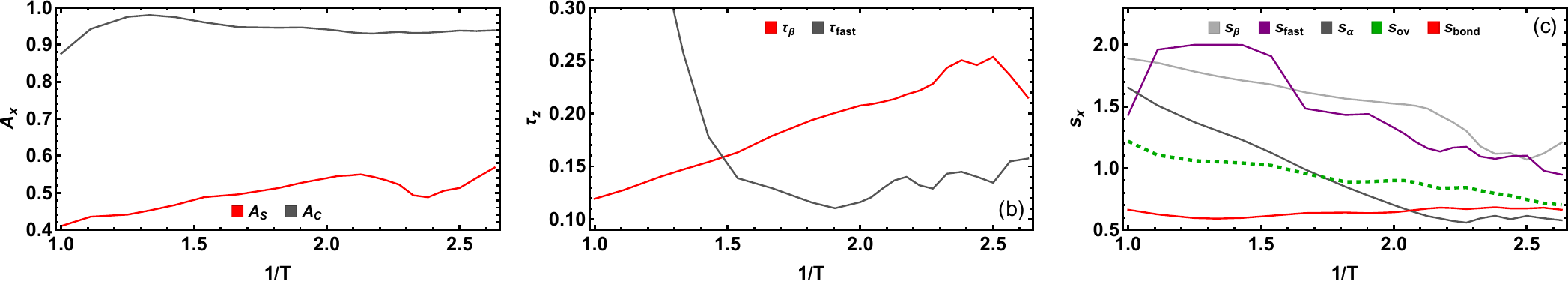}
\caption{ $A_{\rm S}$, $A_{\rm C}$, $\tau_\beta$, $\tau_{\rm fast}$, $s_{\rm fast}$, $s_\beta$, $s_{\rm ov}$, and $s_{\rm slow}$ values obtained by fitting simulation data for $S(q,t)$, $f_{\rm ov}(t)$, and $C_{\rm B}(t)$ to Eqs.\ \ref{eq:cfSQ}-\ref{eq:cfCB}.  Statistical uncertainties for these parameters are comparable to the scatter of the plotted data\rhedit{.}}
\label{fig:app}
\end{figure*}

We showed that  the peak time $\tau_\chi$ for the susceptibility $\chi_{\rm B}(t)$ associated with the spatial heterogeneity defined by $C_{\rm B}(t)$ remains on the order of $\tau_{\rm imm}$ (the characteristic lifetime of immobile-particle clusters), even as both of these quantities varied by roughly $5$ orders of magnitude over our studied range of $T$.
Examining the Van Hove correlation functions evaluated at $t = \tau_\chi(T)$ reveal that diffusion remains strongly non-Gaussian on this timescale -- increasingly so as $T$ decreases -- owing to the persistence of large immobile-particle clusters  over this timescale.

Similarly, we showed that the bond lifetime $\tau_{\rm bond}$ associated with the terminal decay of $C_{\rm B}(t)$ remains on the order of the ``overlap time''  $\tau_{\rm ov}$ over which a typical particle moves by at least half its diameter away from its initial position, even as both of these quantities (much like $\tau_\chi$ and $\tau_{\rm imm}$) varied by roughly 5 orders of magnitude.
A key step in making this connection was our decision to use $a = 1/2$ in Eqs.\ \ref{eq:fovself}-\ref{eq:fovmut}, rather than the smaller value ($a = 0.3$) used in previous studies relating neighbor persistence to heterogeneous dynamics.
Specifically, choosing $a = 1/2$ made $\tau_{\rm ov} \simeq \tau_{\rm bond}$, a relationship which is qualitatively different than the $\tau_{\rm ov} \simeq \tau_\alpha \ll \tau_{\rm bond}$ relationship that is observed when $a = 0.3$ \cite{lacevic03,yamamoto98,shiba12}.

Finally we showed that the ratio $\tau_{\rm bond}/\tau_\alpha$ varies substantially and non-monotonically with $T$, peaking at $T \simeq T_{\rm x}$.
Recall that $\tau_\alpha$ starts to increase sharply, and decoupling of fast and slow relaxation processes occurs, as $T$ drops below the crossover temperature $T_{\rm x}$ \cite{ediger96,ediger00}. 
We believe that $\tau_{\rm bond}/\tau_\alpha$ decreases with decreasing $T$ for $T < T_{\rm x}$ because both timescales are controlled primarily by the immobile-particle clusters.
Followup studies that probe the more-deeply-supercooled regime are needed to test this hypothesis more decisively, but one tentative piece of evidence for it is the fact that $s_{\rm bond} \simeq s_\alpha$ in this temperature regime [Fig.\ \ref{fig:app}(c)].

We conclude by speculating on how one might further intrepret the timescale $\tau_{\rm bond}$.
One intriguing possibility is that it provides an easily-accessible estimate of the ``exchange time'' $\tau_{\rm ex}$ \cite{heuer95,heuer97,starr13,richert15,paeng15}, which is the average time it takes for a “fast” (high-mobility) region to become “slow” (low-mobility) and vice versa.
While $\tau_{\rm ex}$ is conceptually straightforward, mapping it onto a specific observable which can be measured in simulations, let alone experiments, has proven challenging \cite{richert15}.
Traditionally it has been estimated using four-point correlation functions \cite{heuer97,yamamoto98,flenner04,mizuno10,kim13}.
A recently published simulation study used a novel theoretical approach to calculate $\tau_{\rm ex}$ from \textit{two}-point spatiotemporal correlations of local relaxation rates $\gamma(\vec{r},t)$ \cite{pandit23}, but associating $\tau_{\rm ex}$ with $\tau_{\rm bond}$ would provide an even simpler measure of $\tau_{\rm ex}$ that directly associates it with the timescale over which particles forget their local environments.
It would be interesting to attempt to do so.

We thank Ludovic Berthier and Camille Scalliet for helpful discussions.
This material is based upon work supported by the National Science Foundation under Grant Nos.\ DMR-2026271 and DMR-2419261.

\section{Appendix}
\label{sec:appendix}

Additional results obtained from fitting the $S(q,t)$, $f_{\rm ov}(t)$ and $C_{\rm B}(t)$ data to Eqs.\ \ref{eq:cfSQ}-\ref{eq:cfCB}  are shown in Figure \ref{fig:app}.
Panel \rhedit{8a} shows our results for the $A(T)$.
Fig.\ \ref{fig:1} had already made it clear that the $A(T)$ are much larger for $C_{\rm B}(T)$ than they are for $S(q,t)$, but this data shows that they are also both nearly $T$-independent for $T \gtrsim 0.7$.
For larger $T$, these data are less reliable because the nearly-single-step nature of the decays of $S(q,t)$ and $C_{\rm B}(t)$ produces ambiguities in fits of these data to Eqs.\ \ref{eq:cfSQ} and \ref{eq:cfCB}.

Panel \rhedit{8b} shows that the temperature dependence of the ``beta'' relaxation times $\tau_\beta$ and $\tau_{\rm fast}$ is much weaker than it is for slow timescales like $\tau_\alpha$ and $\tau_{\rm bond}$.
This is expected from the well-known weak temperature dependence of $\beta$ relaxation \cite{ediger96}.
In stark contrast to results for the various $\tau_{\rm slow}$, the ratio $\tau_{\rm fast}/\tau_\beta$ remains nearly constant (and of order one) for all $T$ for which the decay of $S(q,t)$ and $C_{\rm B}(t)$ clearly has a two-step character, i.e.\rhedit{,} for all $T \lesssim \rhedit{T_{\rm A}}$.
It would be interesting to investigate this fast decay mode further, as the dynamics of $\beta$ relaxation are quite complex and remain a topic of active study \cite{betancourt18,zhang21,jiang24}.
Note that as mentioned above, in contrast to previous studies \cite{yamamoto98,shiba12,iwashita13,scalliet22}, here we did not suppress the fast decay mode of $C_{\rm B}(t)$ by including a ``padding'' of the distance cutoff for particles to be considered neighbors.

\begin{table*}[htbp]
\caption{Equilibrium densities $\rho = N/V$ and best-fit values for the seven relaxation times discussed in Sec.\ \ref{subsec:IIIB}.  All times are given to two significant figures because the estimated statistical uncertainties on most quantities are of order 1\%.  The uncertainties for $\tau_{\rm mob}$ and $\tau_{\rm imm}$ are larger for $T \leq 0.4$ owing to the limited sampling, while those for $\tau_{\rm mob}$ and $t^*$ are larger for $T \geq 0.9$ where DH is weak.}
\begin{ruledtabular}
\begin{tabular}{lcccccccc}
\hline
$T$ & $\rho$ &  $\tau_{\rm bond}$  & $\tau_\chi$ & $\tau_\alpha$ & $\tau_{\rm ov}$ &  $\tau_{\rm imm}$ & $\tau_{\rm mob}$ & $t^*$ \\
$1$ & $0.933$ & $2.7$ & $0.52$ & $0.24$ & $1.2$ & $1.6$ & $1.2$ & $0.84$ \\
 $0.9$ & $1.016$ & $3.6$ & $0.76$ & $0.29$ & $1.8$ & $2.0$ & $0.88$ & $0.88$ \\
$0.8$ & $1.085$ & $5.5$ & $1.2$ & $0.37$ & $2.6$ & $2.4$ & $1.2$ & $1.0$ \\
$0.75$ & $1.116$ & $7.2$ & $1.6$ & $0.42$ & $3.3$ & $3.2$ & $1.4$ & $1.2$ \\
$0.7$ & $1.147$ & $10$ & $2.2$ & $0.49$ & $4.2$ & $3.8$ & $1.4$ & $1.3$ \\
$0.65$ & $1.177$ & $15$ & $3.2$ & $0.60$ & $5.7$ & $4.6$ & $1.6$ & $1.6$ \\
 $0.6$ & $1.206$ & $24$ & $5.2$ & $0.80$ & $8.8$ & $6.4$ & $2.0$ & $2.2$ \\
 $0.55$ & $1.235$ & $42$ & $9.4$ & $1.2$ & $16$ & $11$ & $3.2$ & $3.2$ \\
 $0.525$ & $1.248$ & $62$ & $14$ & $1.6$ & $24$ & $15$ & $4.4$ & $4.2$ \\
 $0.5$ & $1.263$ & $1.0 \times 10^2$ & $23$ & $2.4$ & $38$ & $24$ & $5.4$ & $6.4$ \\
 $0.49$ & $1.269$ & $1.3 \times 10^2$ & $29$ & $2.9$ & $48$ & $25$ & $7$ & $7.2$ \\
 $0.48$ & $1.274$ & $1.7\times 10^2$ & $38$ & $3.7$ & $63$ & $38$ & $8.4$ & $9.4$ \\
 $0.47$ & $1.280$ & $2.2\times 10^2$ & $51$ & $5.0$ & $87$ & $57$ & $11$ & $13$ \\
$0.46$ & $1.285$ & $3.1\times 10^2$ & $73$ & $7.3$ & $1.3\times 10^2$ & $66$ & $15$ & $15$ \\
 $0.45$ & $1.291$  & $4.7\times 10^2$ & $1.2\times 10^2$ & $12$ & $2.0\times 10^2$ & $1.1\times 10^2$ & $22$ & $23$ \\
 $0.44$ & $1.297$ & $7.5\times 10^2$ & $1.9\times 10^2$ & $21$ & $3.3\times 10^2$ & $1.9\times 10^2$ & $29$ & $34$ \\
 $0.43$ & $1.302$ & $1.3\times 10^3$ & $3.3\times 10^2$ & $49$ & $6.2\times 10^2$ & $3.5\times 10^2$ & $56$ & $60$ \\
$0.42$ & $1.308$ & $2.7\times 10^3$ & $6.9\times 10^2$ & $1.2\times 10^2$ & $1.4\times 10^3$ & $9.9\times 10^2$ & $99$ & $96$ \\
 $0.41$ & $1.314$ & $6.4\times 10^3$ & $1.7\times 10^3$ & $3.4\times 10^2$ & $3.6\times 10^3$ & $2.8\times 10^3$ & $2.7\times 10^2$ & $1.9\times 10^2$ \\
$0.4$ & $1.319$ & $2.0\times 10^4$ & $5.6\times 10^3$ & $1.5\times 10^3$ & $1.3\times 10^4$ & $1.3\times 10^4$ & $8.3\times 10^2$ & $4.6\times 10^2$ \\
$0.39$ & $1.325$ & $7.4\times 10^4$ & $2.2\times 10^4$ & $7.0\times 10^3$ & $5.5\times 10^4$ & $4.4\times 10^4$ & $3.1\times 10^3$ & $9.4\times 10^2$ \\
$0.38$ & $1.330$ & $3.2\times 10^5$ & $9.9\times 10^4$ & $3.7\times 10^4$ & $2.6\times 10^5$ & $3.2\times 10^5$ & $9.7\times 10^3$ & $3.0\times 10^3$ \\
\hline
\end{tabular} 
\end{ruledtabular}
\label{tab:paramvals}
\end{table*}

Panel \rhedit{8c} shows our results for the fast and slow stretching exponents.
$s_{\rm fast}$ tracks and is comparable to the exponent $s_\beta$ over the entire range of $T$, further supporting our identification of the fast decay mode of $C_{\rm B}(t)$ with $\beta$ relaxation.
Note that our results for $s_\beta(T)$ are in qualitative agreement with several previous studies \cite{allegrini99,betancourt18,zhang21,jiang24}.
We also find that $s_{\rm bond}$ is nearly $T$-independent; it remains below $s_{\rm ov}$ for all $T$, and is comparable to $s_\alpha$ for $T \lesssim T_{\rm x}$. $\alpha$ relaxation and neighbor persistence being similarly spatially heterogeneous in this $T$ regime is physically reasonable if both are controlled primarily by the immobile-particle clustsers \cite{starr13,mizuno10}.

 Finally, Table \ref{tab:paramvals} presents numerical values of the seven relaxation times discussed in Sec.\ \ref{subsec:IIIB}.
 Note that these times are likely to be substantially shorter than the corresponding times in the other recent study of the 2:1 KA model's dynamics at these $T$ \cite{ortlieb23} \rhedit{because} that study employed 
  \textit{NVT}-ensemble simulations performed at a fixed density $\rho = N/V = 1.4\sigma_{\rm AA}^{-3}$ which is much larger than any of the densities considered here.
 The KA model's dynamics slow down rapidly with increasing pressure, particularly for low $P$ \cite{yuan24,yuan24b}.


%

\end{document}